\newcommand{\PSD}{\hat{\psi}^\dagger}
\newcommand{\PSI}{\hat{\psi}}
\newcommand{\rv}{({\bf r})}

\documentstyle[pra,aps,twocolumn,epsf]{revtex}
\begin{document}

\draft

\title {
Instability of Non-vortex State toward a
Quantized Vortex in Bose-Einstein Condensate under 
External Rotation
}

\author{
  Tomoya Isoshima\thanks{E-mail: tomoya@mp.okayama-u.ac.jp}
  and Kazushige Machida
}

\address{
  Department of Physics, Okayama University, Okayama 700-8530, Japan
}

\date{\today}

\maketitle

\begin{abstract}
The instability condition of the non-vortex state toward vortex formation is examined within the Bogoliubov theory when a Bose-Einstein condensate is under externally forced rotation.
The obtained critical angular velocity combined with the previous stability conditions for a votex yields a detailed phase diagram in the critical velocity vs the system parameter.
This facilitates vortex formation experiments for alkali atom gases confined in a harmonic potential.
\end{abstract}


The quantized vortex is a manifestation of macroscopic quantum phenomena in quantum fluids  and have been observed in superconductors and superfluid helium.
Since the experimental realization\ \cite{dalfovoReview} of Bose-Einstein Condensation (BEC) in alkali-atom gases, the vortex problem in weakly interacting Bose gas has been investigated theoretically\ \cite{dalfovoReview}.
So far there is no experimental report to find a vortex in the present BEC systems and several trials are ongoing.

It is well known  that the  Gross-Pitaevskii (GP) theory\ \cite{gross,pitaevskii} is valid for describing a weakly interacting Bose system which is the case for the present alkali-atom gases of Rb and Na.
Based on the GP theory the single vortex state is discussed by several authors\ \cite{gpvortex,sinha}.
More recently Butts and Rokhsar\ \cite{ButtsRokhsar} describe a many vortex state in rotating BEC, giving rise to an interesting vortex configuration with many vortices.
This may be directly observable experimentally.
The time evolution of vortex is also studied by a time dependent form of the GP equation\ \cite{gp2dtime}.

To investigate further fundamental issues on a quantized vortex, we resort to the Bogoliubov theory\ \cite{bogoliubov,fetterOlder}.
This microscopic framework can describe the condensate and the non-condensate components simultaneously, yielding the collective excitation spectrum by solving an eigenvalue equation\ \cite{isoshima1,isoshima2,omegaOne,dodd,rokhsar,svi1}.

In our series of papers\ \cite{isoshima1,isoshima2,omegaOne}, we have been studying  the stability problem of an isolated singly quantized vortex.
The main conclusions drawn from our studies are:
(1)
Contrary to the conclusion in classical papers\ \cite{pitaevskii,fetterOlder},  BEC cannot sustain the stable vortex in an infinite system at $T$=0\ \cite{isoshima1}.
This conclusion coincides with Rokhsar\ \cite{rokhsar}.
(2)
The finite temperature which increases the non-condensate component relative to the condensate one effectively stabilizes it in a harmonically confined system\ \cite{isoshima2}. 
(3) 
A pining potential added to a harmonic confined potential also stabilizes the quantized vortex\ \cite{isoshima2}.
We note that concerning the conclusion (2), Stringari\ \cite{stringaryGlobalT} gives the phase diagram of the stability in a wide range of $T$ by comparing total free energies.


In \cite{omegaOne}, apart from the above intrinsic vortex stability problem, we have discussed the vortex stability under externally forced rotation at $T=0$.
There we introduce two different criteria for stability:
The global stability is defined by comparing the two total energies of the non-vortex state and the vortex state.
The other local stability is defined by the intrinsic stability condition for the vortex itself, evidenced by the positiveness of the lowest eigenvalue.
According to these criteria, we obtain the phase diagram (see Fig.7 in Ref.\ \cite{omegaOne}) in the rotation angular frequency $\Omega$ vs the system parameter $an_z$ ($a$: the scattering length and $n_z$: the particle density per unit length), which shows the three divided regions upon increasing $\Omega$: (A) the vortex unstable region, (B) the single vortex stable region, and (C) the single vortex unstable region where many vortices may appear.

In order to quantify the problem further and refine this phase diagram, we investigate the local stability of the non-vortex state under external rotation.
Through this study we may gain a valid picture of the actual nucleation process when we begin to rotate a BEC quite slowly from zero, which has not been examined before.
It is our hope that this study fosters the observation of a vortex in the present BEC systems.

Since the formulation and computational procedures are same as before\ \cite{omegaOne}, we briefly introduce the main equations and the notations.
Let us start the following Hamiltonian for interacting Bosons with the s-wave scattering length $a$:
$
  \hat{\text{H}}_{\text{rot}}
  =
  \hat{\text{H}} - {\bf \omega}\cdot\int{\bf r}\times{\bf p}\rv  d {\bf r}
$
where
$
  \hat{\text{H}}=
  \int  d {\bf r}
    \hat{\Psi}^{\dagger}\rv \{
      -\frac{\hbar^2 \nabla^2}{2m} + V\rv - \mu
    \} \hat{\Psi}\rv 
  + \frac{g}{2} 
  \!\! \int \!\!  d {\bf r} 
  \hat{\Psi}^\dagger \rv \hat{\Psi}^\dagger \rv
  \hat{\Psi}\rv \hat{\Psi}\rv          ,
  $
where
${\bf p}(\bf r)$ is the momentum operator,
$\mu$ is the chemical potential, and
$V\rv$ is the confining potential.
The mass of a particle is $m$.

Following the standard method,  we decompose the field operator $\hat{\Psi}$ as $\hat{\Psi}\rv = \PSI\rv + \phi\rv$, $\phi\rv \equiv \langle \hat{\Psi}\rv \rangle$.
We substitute the above decomposition into the above and ignore the higher order terms such as  $\PSD \PSI \PSI$, $\PSD \PSD \PSI$, and $\PSD \PSD \PSI \PSI$ terms.
Then we rewrite the operator $\PSI\rv$ as
$\PSI\rv = \sum_q
\left[  u_q\rv \eta _q - v_q^*\rv \eta _q^{\dagger} \right]$
where $\eta_q$ are the annihilation operators, 
$u_q\rv$ and $v_q\rv$ are the wave functions,
and the subscript $q$ denotes the quantum number.

We consider a cylindrically symmetric system and a vortex line, if exists,  passes through the center of a cylinder, coinciding with the rotation axis.
We use the cylindrical coordinates:  ${\bf r} = (r, \theta , z)$.
The system is trapped radially by a harmonic potential $ V(r) = \frac{1}{2} m(2\pi \nu)^2 r^2 $ and is periodic along the $z$-axis whose length is $L$.

We focus on the lowest eigenstates of the momentum along $z$-axis and the quantum number ${\bf q}$  is written as $(q_r, q_{\theta})$, where $q_r = 1, 2, \cdots $ and  $q_\theta = 0, \pm 1, \cdots$.
The condensate wave function $\phi\rv $ is expressed as $\phi(r,\theta ,z) = \phi(r) \text{e}^{i w\theta }$. ($\phi(r)$:real)
In similar way, 
$u_{q}\rv = u_{q}(r){\rm e}^{i(q_{\theta } + w)\theta}$
$v_{q}\rv = v_{q}(r){\rm e}^{i(q_{\theta } - w)\theta }.$
The case $w=0$ corresponds to a system without a vortex and $w=1(w=2)$ with a singly  (doubly) quantized vortex.

The condensate has $N$ particles in the length $L$ and the area density is $n_{z} = N/L$.
We introduce the density unit $n_{0} \equiv \sqrt{h\nu /g}$ where $g = 4\pi \hbar^{2}a/m$, the length unit $\sqrt{2mh\nu}\, /\hbar$.
The energy ($\varepsilon^\prime,\mu^\prime$) are scaled by $h\nu$.
The wavefunctions ($\phi, u, v$) are scaled by $\sqrt{n_{0}}$.
The  angular velocity $\omega$ is also scaled by $2\pi\nu$ and the normalized rotation is $\Omega \equiv \frac{\omega}{2\pi\nu}$.
We use the normalized values of $\mu, \varepsilon, \phi, u, v, r, \text{ and } \Omega$ below.

The condition that the first order term in $\PSI\rv$  of our Hamiltonian vanish is  the Gross-Pitaevskii equation
%
\begin{eqnarray}
  \Bigl[
    - \left\{
      \frac{d^2}{dr^{2}}
      +\frac{1}{r} \frac{d}{dr}
      - \frac{w^2}{r^2}
    \right\} - \mu \nonumber
&&\\
    + \frac{r^2}{4} + \phi^{2}(r) + w\Omega 
  \Bigr]\phi(r)
  &=& 0.             \label{eq:gp}
\end{eqnarray}
%
The normalize condition is $\int |\phi(r)|^{2}r\,dr = \frac{4 a N}{L}$.
The condition that the Hamiltonian be diagonalized gives  the coupled eigenvalue equations for $u_q\rv$, and $v_q\rv$ whose eigenvalue is $\varepsilon_q$:
\begin{eqnarray}
  \Bigl[
    -\left\{
      \frac{d^2}{dr^2} + \frac{1}{r} \frac{d}{dr}
      -\frac{(w + q_\theta)^2}{r^2}
    \right\} - \mu          \nonumber
&&\\
    + \frac{r^2}{4} + 2\phi^2(r)
    + \Omega (w + q_\theta )
  \Bigr]u_{q}(r)            \nonumber
&&\\
  - \phi^2(r) v_{q}(r) 
  &=&
  \varepsilon _{q} u _{q} (r) ,    \label{eq:bg1}
\\
  \Bigl[
    - \left\{
      \frac{d^2}{dr^2} + \frac{1}{r} \frac{d}{dr}
      - \frac{(w - q_\theta)^2}{r^2}
    \right\}  - \mu          \nonumber
&&\\
    + \frac{r^2}{4} + 2\phi^2(r)
    + \Omega (w - q_\theta)
  \Bigr] v_{q}(r)          \nonumber
&&\\
  - \phi^2(r) u_{q}(r) 
  &=&
  -\varepsilon _{q} v_{q}(r) .   \label{eq:bg2}
\end{eqnarray}
The normalization conditions are
  $
  \int \left\{
    u_p^{*}(r) u_q(r) - v_p^{*}(r) v_q(r) 
  \right\}r\,dr
  =
  \frac{4 a}{L}\delta_{p,q}.
  $
We determine the signs of $q_{\theta}$ and $\varepsilon$ by the above normalization condition.
The set of the Eqs.\ (\ref{eq:gp}), (\ref{eq:bg1}), and (\ref{eq:bg2}) constitute the Bogoliubov theory of our system.

The contributions of the non-condensate to the total energy is negligibly small and we will use the contribution from the condensate
$
  \frac{E_{\phi}}{h\nu N}=E_{\phi}{N}   
  \frac{1}{4a n_{z}}
  \int [ - \phi^{*}(r)
    (
      \frac{d^2}{dr^{2}}
      +\frac{1}{r} \frac{d}{dr}
      -\frac{w^2}{r^{2}}
    ) \! \phi(r)         
    + \frac{r^2}{4}|\phi(r)|^2
    + |\phi(r)|^4
  ] r \, dr    
  - w \Omega
$
as the total energy of the system.
Here $4a n_{z}$ is a dimensionless number.
Note that the coefficient of $\omega$ in $E_{\phi}$ is a constant.
Since the parameter $\frac{4a}{L}$ in the conditions of normalization does not change the following  results within the present Bogoliubov framework, the system is characterized by the number $a n_{z}$ and the rotation $\Omega$.

\begin{figure}
  \begin{center}
    \leavevmode
    \epsfxsize=8cm
    \epsfbox{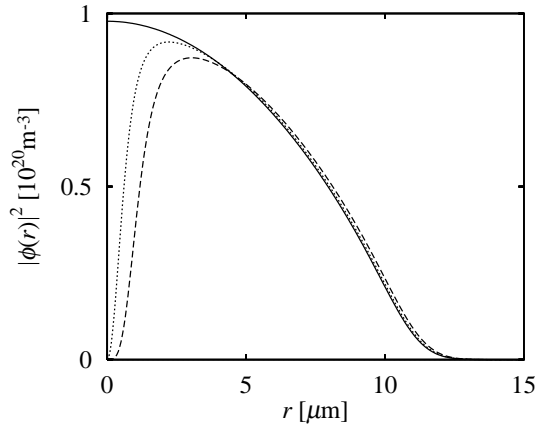}
  \end{center}
  \caption{
    The particle number density distribution $\phi^2(r)$ along the radial direction. 
    Dashed (dotted) line is the singly (doubly) quantized  vortex case $w=1$ ($w=2$) while   solid line corresponds to the  non-vortex case.
    To draw this figure in actual units, we use the following parameters:  the scattering length $a = 2.75 \text{nm}$,   $m = 3.81 \times 10^{-26} \text{ kg}$,   the radial trapping frequency $\nu = 100 \text{Hz}$,  and $n_z = 2 \times 10^4 /\mu \text{m}$.
  }
  \label{fig:dns}
\end{figure}

We solve the coupled equations; Eqs.\ (\ref{eq:gp}), (\ref{eq:bg1}), and (\ref{eq:bg2}) for the gas of $^{23}\text{Na}$ atoms trapped radially by a harmonic potential of $\nu \text{Hz}$.
The area density per unit length along the $z$-axis is chosen to be $a n_z = 2.75 \sim 137.5$.
The density profile of the condensate $\phi^2(r)$ is solved from the GP equation Eq.\ (\ref{eq:gp}).
Figure\ \ref{fig:dns} shows the profiles of the system with $w=0,1,\text{ and } 2$.

The eigenvalues $\varepsilon$'s evaluated from  Eqs.\ (\ref{eq:bg1}) and (\ref{eq:bg2}) should be all positive relative to the condensate energy at zero for the system to be stable.
This defines the local stability of the system.
In \cite{omegaOne}, we have derived the critical $\Omega$'s of the local stability in vortex state.
The lower criterion is $\Omega_{\text{local}}^{\text{L}}$ and the upper criterion is $\Omega_{\text{local}}^{\text{U}}$.
For example,  $\Omega_{\text{local}}^{\text{L}} =0.16085$ and $\Omega_{\text{local}}^{\text{U}} =0.44610$ when $a n_{z}=55$.

Also the non-vortex state, which has the winding number $w=0$, could become locally unstable under the large external rotation.
In other words, some excitation level $\varepsilon_{\bf q}$ at
larger $q_{\theta}$ can become negative.
These negative excitations correspond to ``surface excitations'' 
derived by Dalfovo\ \cite{dalfovoSurface}.

\begin{figure}
  \begin{center}
    \leavevmode 
    \epsfxsize=8cm
    \epsfbox{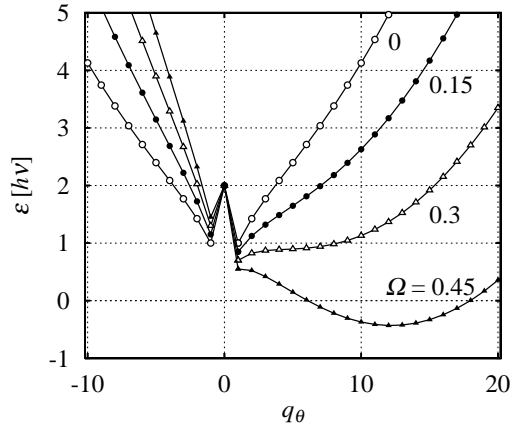}
  \end{center}
  \caption{
    The lowest edge of the eigenvalues along $q_\theta$ for selected  angular velocities $\Omega$'s.
    Circles, filled circles, triangles, and filled triangles   correspond  to $\Omega$ = 0, 0.15, 0.3, and 0.45 respectively ($an_z=55$).
    Another eigenvalues is densely distributed at each $q_{\theta}$.
    The eigenvalues in $q_\theta<0$ move up while the eigenvalues  in $q_\theta>0$ move down as $\Omega$ increases.
    At larger $\Omega$  some positive eigenvalue of $q_{\theta}$ around $q_{\theta}\sim 10$  becomes negative as $\Omega$ increases.
    The eigenvalue at $q_\theta=0$ does not change with $\Omega$.
  }
  \label{fig:zerowing}
\end{figure}

\begin{figure}
    \leavevmode
    \epsfxsize=8cm
    \epsfbox{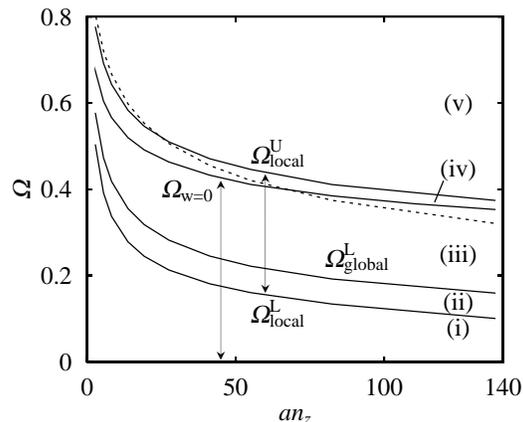}
  \caption{
    The four main critical velocities $\Omega$'s;
    $\Omega^{\text{L}}_{\text{local}}$, 
    $\Omega^{\text{L}}_{\text{global}}$,
    $\Omega_{w=0}$, and
    $\Omega^{\text{U}}_{\text{local}}$
    are plotted in solid lines as functions of $an_z$.
    These four lines divide the whole area into five regions (i) - (v).
    Each of regions are explained in text and in Fig.\ \protect\ref{fig:trans}.
    The dashed line denotes $\Omega^{\text{U}}_{\text{global}}$.
    The normalization condition for Eq.\ (\protect\ref{eq:gp});$\int |\phi(r)|^{2}r\,dr = 4a n_z$ will be helpful to understand the value $an_z$. 
  }
  \label{fig:LandG}
\end{figure}

Here, we calculate the critical $\Omega$ in the non-vortex state to refine the phase diagram (Fig.\ 7 in \cite{omegaOne}) of stabilities.
In Fig.\ \ref{fig:zerowing} the lowest edge of the eigenvalues in the non-vortex state is   displayed for selected $\Omega$'s.
There is a critical value of $\Omega_{w=0}$ above which some eigenvalue(s) in $q_{\theta}>0$ become(s) negative, signalling an instability of the non-vortex state.
When $\Omega<\Omega_{w=0}$ all the eigenvalues are positive.
When $a n_{z}=55$, $\Omega_{w=0} = 0.41145$.
The resulting curve for $\Omega_{w=0}$ is drawn as a function of $a n_z$ in Fig.~\ref{fig:LandG}.



Besides the above local stability conditions for the non-vortex and vortex states, there is the global stability condition for $\Omega$. The comparison of the two energies $E_{\phi}$ in the systems with and without a vortex defines the ``global stability''.
The critical value between non-vortex state and single vortex state is $\Omega^{\text{L}}_{\text{global}}$ and that between single vortex state ($w=1$) and double vortex state  ($w=2$) is $\Omega^{\text{U}}_{\text{global}}$.
The critical value is $\Omega^{\text{L}}_{\text{global}} = 0.22215$ and $\Omega^{\text{U}}_{\text{global}} = 0.42071$ when $a n_{z} = 55$.

\begin{figure}
    \leavevmode
    \epsfxsize=8.4cm
    \epsfbox{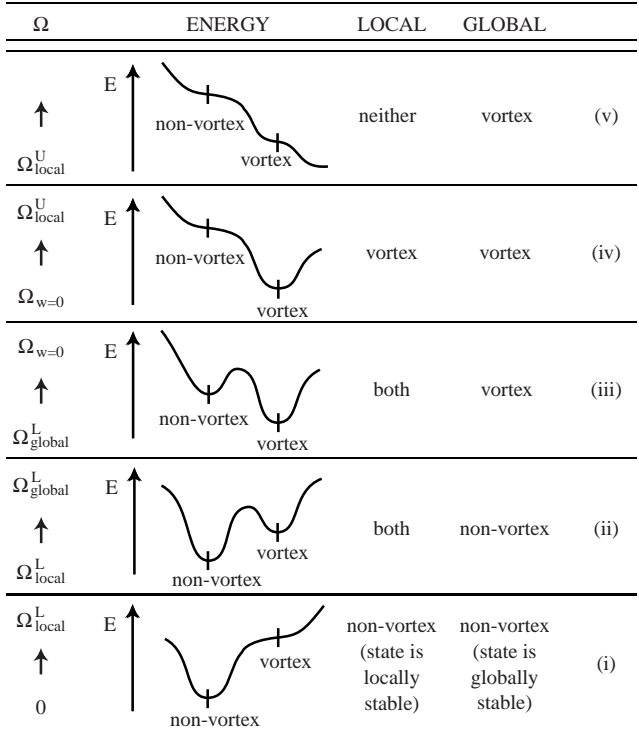}
  \caption{
    From left to right column, the ranges of $\Omega$, 
    the schematic energy configuration near non-vortex and    vortex states,
    the local stability, and the global stability.
    The number of the last column corresponds to that of
    Fig.~\protect\ref{fig:LandG}.
  }
  \label{fig:trans}
\end{figure}

As mentioned previously,  the system is characterized by the number $a n_{z}$ and the rotation $\Omega$.
In Fig.~\ref{fig:LandG} we summarize the four critical $\Omega$'s; $\Omega^{\text{L}}_{\text{local}}$,  $\Omega^{\text{L}}_{\text{global}}$, $\Omega_{w=0}$, and $\Omega^{\text{U}}_{\text{local}}$ as a function of $a n_{z}$.
We can divide the whole range of $\Omega$ into five regions, which are labeled as (i) to (v) in Fig.~\ref{fig:LandG}.
Each region is characterized in Fig.\ \ref{fig:trans} where the schematic energy curve for each region is shown.

As $\Omega$ increases from zero,
\\
(i) the non-vortex state is stable both locally and globally for $0<\Omega<\Omega^{\text{L}}_{\text{local}}$.
\\
(ii) The singly quantized vortex also becomes stable locally for $\Omega^{\text{L}}_{\text{local}}<\Omega<\Omega^{\text{L}}_{\text{global}}$.
But the vortex state has higher energy than that in the non-vortex state.
Therefore, a vortex does not nucleate spontaneously in this region.
\\
(iii) The energy of this vortex state becomes lower than that in the corresponding non-vortex state for $\Omega^{\text{L}}_{\text{global}}<\Omega<\Omega_{w=0}$.
As it is seen from Fig.\ \ref{fig:trans}, the vortex state is locally stable now, but the non-vortex state is still locally stable.
Therefore, the non-vortex state must tunnel quantum-mechanically through an energy barrier to the vortex state.
\\
(iv) The vortex state can nucleate spontaneously in $\Omega_{w=0}<\Omega<\Omega^{\text{L}}_{\text{local}}$, because the non-vortex state is unstable both locally and globally, and there is no reason for the system to stay in the non-vortex state anymore.
A vortex is now bound to appear. 
\\
(v) Since single vortex state becomes unstable, many vortices may appear in this region: $\Omega^{\text{L}}_{\text{local}}<\Omega$.

It is to be noted that even above the $\Omega_{\text{global}}^{\text{L}}$ which is usually regarded as a critical angular velocity of vortex formation, the non-vortex state remains locally stable in region (iii).
The upper edge of (iii) region ($\Omega_{w=0}$)  exceeds twice of $\Omega_{\text{global}}^{\text{L}}$ at larger $a n_{z}$.
This means that the angular velocity much larger than $\Omega_{\text{global}}^{\text{L}}$ is needed to create a vortex.

The comparison between $E_{\phi,w=2}$ and $E_{\phi,w=1}$ defines $\Omega^{\text{U}}_{\text{global}}$.
The energy $E_{\phi,w=2}$ may become lower than that of the  state with two singly quantized vortices (this energy cannot be calculated in this axial symmetry formulation).
But in the discussion below, we treat $\Omega^{\text{U}}_{\text{global}}$ as the (tentative) angular velocity to stabilize many vortices states.

We have ignored $\Omega^{\text{U}}_{\text{global}}$ in Figs.\ \ref{fig:LandG} and \ref{fig:trans}.
This does not contradict with the above explanations in (iii) or (iv) region.
But at much smaller $an_{z}$,  the $\Omega^{\text{U}}_{\text{global}}$ appears in (v) region.
It means that when the  $w=1$ state is unstable locally toward many vortices state, the energy of $w=2$ vortex state is higher than that of the $w=1$ state.
It seems to contradict with the illustration in the first line of Fig.\ \ref{fig:trans}.
A possible explanation is as follows: 
The radius of $w=2$ vortex is so large that it is comparable with the system's radius with $w=1$.
This increases $E_{w=2}$ and thus $\Omega^{\text{U}}_{\text{global}}$.

In order to obtain more insights into the vortex stability problem in the Bose-Einstein condensation of alkali atom gases confined in a harmonic potential, we have extended our previous work~\cite{isoshima1,isoshima2} to the case where  the cylindrical system is under forced rotation.

Our calculation is done within the framework of the non-selfconsistent Bogoliubov theory and yielded the vortex stability phase diagram shown in Fig.~\ref{fig:LandG}.
The stability of the vortex state is examined by the two different ways. They are the local stability and the global stability.
The local stability is evaluated both in single vortex state and in non-vortex state.
These stability conditions yield the whole phase diagram of the vortex stability problem shown in Fig.~\ref{fig:trans}.
In addition to the usual critical angular velocity $\Omega_{\text{global}}$, we quantitatively obtain the values of $\Omega_{w=0}$ and $\Omega_{\text{global}}$ which describes the instability of the non-vortex state toward the vortex state.

The previous cases~\cite{isoshima1,isoshima2}, which are consistent with the present results, correspond to the $\Omega=0$ axis in Fig.~\protect\ref{fig:LandG} where the vortex is  intrinsically unstable.
This instability is seen to occupy a finite range of $\Omega$, not only the isolated line confined at $\Omega =0$, for various densities $n_z$ and the scattering lengths $a$ as another axis.
It is found that the angular velocity much larger than $\Omega_{\text{global}}^{\text{L}}$  is needed to create a vortex.
With the previous finite temperature calculations~\cite{isoshima2}, the present calculation concludes that alkali atom Bose gases in BEC can sustain and  exhibit the stable vortex in certain range of temperature and external rotation.

There are many problems left concerning the vortices in BEC.
About the appropriate value of angular velocity, we need the calculations in full three dimension geometry.
The process to gain/lost the vortex core will be described better if we can treat the time evolution of both the condensate and the non-condensate\ \cite{nikuniTwoCompo}.
The way to give the angular momentum to the condensate (or cloud of atoms before condensation) is the important problem for experimentalists \ \cite{laserkarl}.
It is our hope that this study fosters the observation of a vortex in the present BEC systems.

The authors thank L. Pitaevskii for raising an adequate and timely question regarding to the instability of the non-vortex state, which leads to the present work.



\end{document}